\documentclass[%
 reprint,
 amsmath,amssymb,
 aps,
]{revtex4-2}
\usepackage[a4paper,margin=1in]{geometry}
\usepackage{tikz}
\usepackage{graphicx,float}
\usepackage{quantikz}
\usepackage{caption}
\usepackage[utf8]{inputenc}
\usepackage{subcaption}
\usepackage{dcolumn}
\usepackage{bm}
\usepackage[ruled,linesnumbered]{algorithm2e}
\usepackage{amsmath,amsfonts,amssymb,mathtools}
\usepackage{multirow}
\usepackage{eso-pic}
\usetikzlibrary{arrows.meta}
\usetikzlibrary{shapes.geometric, arrows}
\usepackage{comment}
\usepackage[colorlinks=true, linkcolor=cyan, urlcolor=cyan, citecolor=cyan]{hyperref}
\usepackage{url} 

\begin{document}

\preprint{APS/123-QED}

\title{ A hybrid classical-quantum approach to highly constrained \\ Unit Commitment problems}
%

\author{Bruna Salgado$^{1,2}$} \email{bruna.f.salgado@inesctec.pt}
\author{André Sequeira$^{1,2,3}$} 
\author{Luis Paulo Santos$^{1,2,3}$} 

\affiliation{$^1$ Department of Informatics, University of Minho}%
\affiliation{$^2$ HASLab, INESC TEC}%
\affiliation{$^3$ International Nanotechnology Laboratory (INL)\\ Braga, Portugal \\}

\date{\today}

\begin{abstract}
The unit commitment (UC) problem stands as a critical optimization challenge in the electrical power industry. It is classified as NP-hard, placing it among the most intractable problems to solve. This paper introduces a novel hybrid quantum-classical algorithm designed to efficiently (approximately) solve the UC problem in polynomial time.
In this approach, the UC problem is decomposed into two subproblems: a QUBO (Quadratic Unconstrained Binary Optimization) problem and a quadratic optimization problem. The algorithm employs the Quantum Approximate Optimization Algorithm (QAOA) to identify the optimal unit combination and classical methods to determine individual unit powers.
The proposed hybrid algorithm is the first  to include both the spinning reserve constraint  (thus improving its applicability to real-world scenarios) and to explore QAOA warm-start optimization in this context. The effectiveness of this optimization was illustrated for specific instances of the UC problem, not only in terms of solution accuracy but also by reducing the number of iterations required for QAOA convergence.
Hybrid solutions achieved using a single-layer warm-start QAOA ($p=1$) are within a 5.1\% margin of the reference (approximate) classical solution, while guaranteeing polynomial time complexity on the number of power generation units and time intervals. 
\end{abstract}

\maketitle


\section{Introduction}

Unit Commitment (UC) is one of the most important optimization problems in the electrical power industry. It aims to determine the optimum scheduling of power generating units \textit{i.e.} determining the on/off status, as well as the power generation levels that minimize operational cost over a certain time horizon. Optimally scheduling power generation units can lead to substantial savings, potentially amounting to millions of dollars, by reducing fuel consumption and associated operational costs \cite{padhy2004unit}.

A most simplified version of the UC problem can be expressed as
\begin{subequations}
\begin{align} 
 \min\limits_{\substack{p_{1 \ldots N,1 \ldots T} \\ y_{1 \dots N,1 \ldots T}}} \quad & P = \sum_{t=1}^{T} \sum_{i=1}^{N} \phi_{i, t}(p_{i,t}) \hspace{2pt} y_{i,t} \nonumber  \\
 s.t. \quad  & \sum_{i=1}^{N} p_{i,t} \hspace{2pt} y_{i,t} = L_t \nonumber \\ 
 & p_{i}^{min} \leq p_{i,t} \leq p_{i}^{max} \nonumber 
 \end{align}    
\end{subequations}

where $y_{i,t}$ and $p_{i,t}$ indicate, at time
$t$, whether unit $i$ is committed or
not and the power it generates, respectively. $P$ is the production cost, $\phi_{i, t}(p_{i,t})$ is the, possibly non-linear, cost function, and  $L_t$ is the total power load.  $p_{i}^{min}$ and $p_{i}^{max}$ are the minimum and maximum power generated by unit $i$. In general, the set of constraints is much larger, including system-wide restrictions, such as load balance and spinning reserve, and specific unit constraints, such as unit generation capacity, ramp rate limits, minimum up/down time requirements, and cold/hot start times. These are addressed by this paper and described in detail in Section \ref{sec:uc}.

As current energy systems grow in size and complexity, improving the efficiency of solving UC becomes more important to the industry \cite{ajagekar2019quantum,datta2013unit}. Due to the nonlinear cost function and the combinatorial nature of the set of feasible solutions, this problem is particularly difficult to solve. It has been proven that UC is  NP-hard, which means that it is considered to be in the class of most intractable problems.  \cite{tseng1996power,1270468}

The Quantum Approximate Optimization Algorithm (QAOA) \cite{farhi2014quantum} is a hybrid quantum-classical algorithm that can be used to find approximate solutions to combinatorial optimization problems in polynomial time \cite{blekos2024review}. In particular, if a combinatorial problem can be expressed as a quadratic unconstrained binary optimization (QUBO) problem, then it can be approximately solved using QAOA. 

The canonical QUBO framework features binary choices and does not include constraints. Nevertheless, the framework can be extended by integrating constraints into the objective function through the means of slack variables and indicator variables. While it is possible to address continuous variables by discretizing  their value intervals, this would significantly increase the required number of qubits, in proportion to the number of discrete values considered. Given the limitations of current quantum machines, this is not an ideal solution. An alternative approach for integrating non-binary decisions is to decompose the UC formulation into a QUBO and a quadratic subproblem, and employ a classical optimizer to handle the latter. 

As noted in the previous paragraph, the constrained binary choice part of UC formulation can be expressed as a QUBO. This makes QAOA a compelling algorithm to address (the QUBO subproblem of) UC problems, with the potential to generate approximate solutions in polynomial time. This paper follows this approach for highly constrained UC problems. 

QAOA evolves an initial state to some final state through a quantum circuit, with a state being some assignment to the binary decision variables. It has been shown that an appropriate selection of the initial state significantly increases QAOA's convergence \cite{Egger_2021}. This is known as warm-starting and is particularly advantageous at low depth, which is especially important in the Noisy Intermediate Scale Quantum (NISQ) era.

Warm-starting leverages the solution of a relaxed QUBO problem as the initial state for QAOA. A relaxed solution is an approximate solution that does not fully satisfy all the constraints of the original optimization problem; it loosens some or all of the constraints to make the problem more manageable and simpler to solve.
In the case of QAOA, the solution to the continuous relaxation of the problem is used.

This paper introduces a novel hybrid (quantum plus classical) algorithm for approximately solving highly constrained UC problems in polynomial time. QAOA is used to solve for the binary variables ($y_{i,t}$) and a classical optimizer is used to solve for the non-binary quantities ($p_{i,t}$), within an iterative process. The advantages of wisely warm-starting QAOA are also empirically verified. The proposed approach is demonstrated for a wide range of constraints, including the spinning reserve capacity, which has not been previously addressed in hybrid environments.

The accuracy of the proposed algorithm is demonstrated to be comparable to the best-known classical solutions, with the obtained warm-start results falling within a 5.1\% margin of the reference approximate classical solution. Additionally, the devised algorithm exhibits worst-case polynomial time complexity with respect to the problem size, in contrast to the exponential worst-case time complexity of the reference classical solver.

In Section \ref{sec:Related_Work}, previous research on utilizing QAOA for solving the UC problem is presented and discussed. In Section \ref{sec:uc}, the UC problem is introduced and decomposed into a QUBO and a quadratic  problem. The former is  suitable for employing QAOA, the quantum combinatorial algorithm that is described with some detail in Section \ref{sec:QAOA}. 
Subsequently, Section \ref{sec:Algorithm_Implemented} describes the proposed hybrid algorithm, and Section \ref{sec:Methodology} outlines the methodology. Finally, Section \ref{sec:Results} presents the experimental results obtained with a classical quantum simulator.

\section{Related Work} \label{sec:Related_Work}

Ajagekar et al. \cite{ajagekar2019quantum} reformulate the UC problem as a QUBO by discretizing the problem space, \textit{i.e.}, segmenting the range of values between the minimum and maximum power generation capacities into $n$ discrete values. 
This reformulation increases the number of binary variables from $N$ to $nN$, where $N$ represents the total number of units. A quantum annealing algorithm is then employed to solve the resulting combinatorial problem. The results indicate that this method is effective for small problem sizes, although the algorithm's performance deteriorates as the number of units increases. The authors constrained only the load and the minimum and maximum power generation per unit, considering a single time instant.

The authors in \cite{koretsky2021adapting} propose a hybrid algorithm in which the optimal combination of units is determined by QAOA, with subsequent utilization of a classical optimizer to simultaneously optimize the QAOA parameters and power assignments for each unit. Results are presented for up to 10 power units, constraining only the load and the minimum and maximum power generation per unit, and considering a single time instant.

To improve the scalability of QAOA, \cite{nikmehr2022quantum} introduces a decomposition and coordination framework. The central concept of this framework involves breaking down large-scale UC problems into smaller subproblems, which are then modeled in QUBO form and solved using QAOA. The power generated by each subproblem is shared to update the collective variables. The results achieved demonstrate comparability with their respective classical counterparts. In addition to the above-described basic constraints, the following are also taken into account: ramp-up, ramp-down, minimum uptime, minimum downtime constraints, and line power flow constraints. Results are presented for  9-unit systems across 24 times instants.

Along the same lines, \cite{mahroo2022hybrid} decomposes the UC problem into a quadratic subproblem, a quadratic unconstrained binary optimization (QUBO) subproblem, and a quadratic unconstrained  subproblem. The first and third subproblems are solved by a classical optimization solver, while the QUBO subproblem is solved using QAOA. Only the constraints regarding load and the minimum and maximum power generation capacities for each unit were considered. Results are presented for a 10-unit system, considering a single time instant.

Feng et al. \cite{feng2022novel} integrate the QAOA method into the surrogate Lagrangian relaxation (SLR) technique. It decomposes the UC problem into binary subproblems for time-unit splitting and continuous subproblems, solved via quantum and classical computing methods, respectively. This analysis includes a numerical example featuring a 3-unit system across 4 time instants and considers system demand, generation capacity, and ramp-rate constraints.

The authors in \cite{stein2023combining} employ the HHL algorithm within a QAOA framework to address the UC problem, claiming a cubic speedup over classical algorithms. Their analysis factors in time and transmission costs, and their practical example accounts for three units and two-time instants.

This work extends previous results by exploring the warm-start optimization, arguing that the proposed algorithm exhibits a polynomial time complexity and supports a wide range of constraints, including spinning reserve.

\section{Unit Commitment Problem } \label{sec:uc}

The UC problem entails minimizing the production cost of a system comprising N units over a specific time horizon T. This goal is contingent on unit-specific constraints like generation capacity and system-wide constraints, such as overall system demand. It aims to determine which and when each unit should be turned on/off and at which power level the active units should operate.

\subsection{Standard Formulation}

Each unit can be either turned on or off, represented by the binary variable $y_{i,t} \in \{0, 1\}$. When $y_{i,t} = 1$, the power generated by unit $i$ at time instant $t$, denoted as $p_{i,t}$, is a continuous real value with the constraint $p^{\min}_{i} \leq p_{i,t} \leq p^{\max}_{i}$, where $[p^{\min}_i, \ldots, p^{\max}_i]$ is the absolute, time-independent, power operating range of unit $i$. 

The total power generated by all active units at instant $t$ must satisfy the system demand ($L_{t}$) -- equation \ref{eq:unc_b}.

The ramp rate constraint limits the extent of rapid fluctuations in power generation by a unit across consecutive time intervals. Specifically, the ramp-down ($r^{dn}_{i}$) and ramp-up ($r^{up}_{i}$) limits of each unit $i$ define the maximum permissible decrease or increase in the power output of the unit, respectively, at each time instant. The minimum and maximum power output by each unit $i$ at instant $t$, $p^{\min}_{i,t}$ and $p^{\max}_{i,t}$ are determined by the ramp rate constraints, as follows:
\begin{align}
p^{min}_{i,t} =
\begin{cases}
    \max \{p^{\min}_i, p_{i,t-1} - r^{dn}_{i}\}, \\ \hspace{50pt} \text{if }  y_{i,t-1}= y_{i,t}=1 \\
  p^{\min}_i,   \hspace{25pt} \text{otherwise}. 
\end{cases} \label{eq:rampDown}
\end{align}
\begin{align}
p^{max}_{i,t} =
\begin{cases}
    \min \{p^{max}_i, p_{i,t-1} + r^{up}_{i}\}, \\
    \hspace{50pt} \text{if }  y_{i,t-1}= y_{i,t}=1 \\
  p^{max}_i, \hspace{25pt} \text{otherwise}.
\end{cases} \label{eq:rampUp}
\end{align}

The ramp rate constraints lead to the operating range constraint per time interval --- equation \ref{eq:unc_c}.

Furthermore, there exist limitations on the minimum duration a unit must remain continuously `off'  before it becomes eligible for activation, denoted as $T^{down}_{i}$, as well as on the minimum consecutive time the unit must be `on' before it can be turned off represented by $T^{up}_{i}$. These constraints are captured in equations \ref{eq:unc_d} and \ref{eq:unc_e}. The continuously off/on time of unit $i$ up to time $t$, represented by  $t^{\text{off}}_{i,t}$/ $t^{\text{on}}_{i,t}$, respectively,  are determined as follows:

\begin{align}
t^{\text{off}}_{i,t} = 
  \begin{cases}
   0, &\text{ if } y_{i,t}=1    \\
    1, &\text{ if }   y_{i,t}=0, t=1 \\ 
   1 + t^{\text{off}}_{i,t-1}, &\text{ if }   y_{i,t}=0, t>1 
  \end{cases}
\end{align}

\begin{align}
t^{\text{on}}_{i,t} = 
  \begin{cases}
   0, &\text{ if } y_{i,t}=0    \\
    1, &\text{ if }   y_{i,t}=1, t=1\\ 
   1 + t^{\text{off}}_{i,t-1}, &\text{ if }   y_{i,t}=1, t>1 
  \end{cases}
\end{align}

Lastly, the aggregate of maximum power generating capacities from all committed units at a given time instant should equal or exceed the sum of the known power demand and the minimum spinning reserve requirement for that specific time instant ($R_{t}$) --- equation \ref{eq:unc_f}.

Each unit’s operating fuel cost function is defined by three coefficients: $A$ represents the fixed cost per time step incurred by the unit being on, regardless of the power it contributes; $B$ and $C$ are the linear and quadratic coefficients, respectively, and contribute to the unit’s cost based on its power level.  This cost can be expressed as:
\begin{equation}
    \phi_{i,t} = A_{i} +B_{i} p_{i,t} +C_{i} p_{i,t}^{2}
\end{equation}
Additionally to the operating fuel costs of the committed units, the production cost, $P$, also includes the start-up costs
of the uncommitted units. This start-up cost $\psi_{i,t}$ is defined as:
\begin{align}
\psi_{i,t}=
  \begin{cases}
    d_{i}, &\text{ if }  t^{\text{off}}_{i,t} \leq T^{down}_{i} + f_{i}  \\
    e_{i}, &\text{ if }  t^{\text{off}}_{i,t} > T^{down}_{i} + f_{i}
  \end{cases}
\end{align}

Where $d_i$, $e_i$, $f_i$ are the  hot start cost,
cold start cost, cold start time of unit $i$, respectively.

The UC optimization process aims to identify the optimal combination of unit statuses and power generation levels within the time horizon while fulfilling the specified constraints.

In general, a UC problem can be expressed as follows:
\begin{subequations} \label{eq:genuc}
\begin{align} 
    \min\limits_{\substack{p_{1 \ldots N,1 \ldots T} \\ y_{1 \dots N,1 \ldots T}}} & \quad P=\sum_{t=1}^{T}\sum_{i=1}^{N}\left[\phi_{i,t} + \psi_{i,t}  (1-y_{i,t-1}) \right] y_{i,t}  \label{eq:unc_a}\\
    s.t. & \quad \sum_{i=1}^{N} p_{i,t} \hspace{2pt}  y_{i,t}=L_{t} \label{eq:unc_b}\\ 
    & \quad p^{min}_{i,t} \leq p_{i,t} \leq p^{max}_{i,t} \label{eq:unc_c} \\
    & \quad t^{on}_{i,t}  \geq T^{up}_{i}, \text{ if }  (y_{i,t}, y_{i,t-1}) = (0, 1)  \label{eq:unc_d}\\
    & \quad t^{\text{off}}_{i,t}  \geq T^{down}_{i}, \text{ if }  (y_{i,t}, y_{i,t-1}) = (1, 0) \label{eq:unc_e} \\
    & \quad \sum_{i=1}^{N} p^{max}_i \hspace{2pt} y_{i,t} \geq L_{t} + R_{t} \label{eq:unc_f} \\
    & \quad p_{i,t} \in \mathbb{R}_{\geq 0}, \quad y_{i,t} \in\{0,1\} \nonumber
\end{align}
\end{subequations}
\vspace{-5pt}
where 
\begin{itemize}
  \item $\mathrm{N}$ is the number of units;

  \item $\mathrm{T}$ is the time horizon;

  \item $y_{i,t}$ indicates whether unit $i$ is committed at time $t$ or not. When $y_{i,t}=1$ the unit is on, and when $y_{i,t}=0$ the unit is off $\left(p_{i,t}=0\right)$.

  \item $p_{i,t}$ is the power generated by unit $i$ at time instant $t$.

  \item $\mathrm{L_{t}}$ is the total power load at time instant $t$.

  \item $p^{min}_{i,t}$ and $p^{max}_{i,t}$ define the power operating range for unit $i$ at time $t$, and are determined by equations \ref{eq:rampDown} and \ref{eq:rampUp}.

  \item $R_{t}$ is minimum spinning reserve requirement for time instant $t$.

  \item $T^{down}_{i}$ / $T^{up}_{i}$ are the minimum duration a unit $i$ must remain continuously `off' (respectively `on') before it can switch state.

  \item $t^{on}_{i,t}$ / $t^{\text{off}}_{i,t}$ represent the continuous `on'/`off' time of unit $i$ up to time $t$.
\end{itemize}

\subsection{Q(UB)O decomposition} \label{subsec:QUBO_formulation}
To decompose the existing formulation into a Quadratic Unconstrained Binary Optimization (QUBO) problem and a Quadratic Optimization problem, certain constraints must be removed. This can be accomplished by altering the objective function to incorporate penalty terms, which increase the value of the objective function when the constraints are not respected. 

The constraints presented  in equation \ref{eq:genuc} include five types of equations/inequations:
\begin{subequations} \label{subeq:constraints}
\begin{align} 
&\sum_i a_i \cdot b_i = X \label{subeq_constr_a} \\
& a_i \leq X_i \label{subeq_constr_b} \\
& a_i \geq X_i \label{subeq_constr_c} \\
& C_i \cdot b_i \geq X_i \cdot b_i \label{subeq_constr_d} \\
&\sum_i C_i \cdot b_i \geq X \label{subeq_constr_e} 
\end{align}
\end{subequations}
 where $b_i$ corresponds to a binary variable, $a_i$ to a continuous-valued variable, and $C_i, X, X_i$ are positive constants. Equation \ref{subeq_constr_a} is associated with the system demand constraint (equation \ref{eq:unc_b}),
 equations \ref{subeq_constr_b} and \ref{subeq_constr_c} include the minimum and maximum power generation constraints and the ramp-rate constraints (equation \ref{eq:unc_c}),
equation \ref{subeq_constr_d} covers the minimum up and down time constraints (equations \ref{eq:unc_d} and \ref{eq:unc_e}) and equation \ref{subeq_constr_e}  addresses the spinning reserve constraint (equation \ref{eq:unc_f}).

Inequation \ref{subeq_constr_d} addresses both constraints \ref{eq:unc_d} and \ref{eq:unc_e}, since these can be rewritten as 
\begin{align*}
      t^{on}_{i,t} \cdot (1-y_{i,t}) \cdot y_{i,t-1} & \geq T^{up}_{i} \cdot (1-y_{i,t}) \cdot y_{i,t-1}  \nonumber \\
    t^{\text{off}}_{i,t} \cdot y_{i,t} \cdot (1-y_{i,t-1}) & \geq T^{down}_{i} \cdot y_{i,t} \cdot (1-y_{i,t-1}) \nonumber
\end{align*}
  
In the quadratic unconstrained formulation, these constraints are added to the objective function as penalty terms. Each penalty term is weighted by a different coefficient $\lambda$, which allows modulating each constraint impact in the final solution. The penalty terms corresponding to the different constraints in equation \ref{subeq:constraints} are, respectively:

\begin{subequations}
\begin{align} 
&\left( \sum_i [b_i \hspace{2 pt} a_i] - X \right)^{2} \label{subeq_qubo_a} \\
& \left(a_i - X_i - s_{1,i,t}\right)^2 \label{subeq_qubo_b} \\
& \left(a_i - X_i + s_{2,i,t}\right)^2 \label{subeq_qubo_c} \\
& \left(C_i \cdot b_i \cdot \text{flg}_{1,i,t}\right)^2 \label{subeq_qubo_d} \\ 
&\left( \sum_i [C_i \cdot b_i ] - X + s_{3,t} \right)^2 \label{subeq_qubo_e} 
\end{align}
\end{subequations}
$s_{1,i,t}$, $s_{2,i,t}$, and $s_{3,t}$ are slack variables, while $\text{flg}_{1,i,t}$ is a indicator variable. 
In penalty terms \ref{subeq_qubo_b} and \ref{subeq_qubo_c}, if the constraints are met, $s_{1,i,t}$ and $s_{2,i,t}$ are set to $a_i - X_i$ and $X_i - a_i$, respectively, and to $0$ otherwise.  
In the expression \ref{subeq_qubo_d}, the indicator binary variable is set to $0$ if the constraint is satisfied and $1$ if it is violated.
Finally, for penalty term \ref{subeq_qubo_e}, $s_{3,t}$ is set to $X - \sum_i [C_i \cdot b_i ]$ when the constraint is satisfied, and to $0$ otherwise. 

Additionally, with respect to the start-up cost $\psi_{i,t}$, within the quadratic unconstrained formulation it corresponds to 
\begin{equation} \label{eq:qubo_start-up_cost}
    \psi_{i,t} = d_{i}\cdot (1-\text{flg}_{2,i,t}) + e_{i}\cdot \text{flg}_{2,i,t} 
\end{equation}
where the indicator variable $ \text{flg}_{2,i,t}$ is set to 1 when  $ t^{\text{off}}_{i,t} > T^{down}_{i} + f_{i}$ and $0$ otherwise.

Decomposing the resulting quadratic unconstrained binary problem into a QUBO subproblem and a quadratic binary problem involves minimizing two objective functions. In the QUBO subproblem, which only allows binary variables, all non-binary variables are treated as constants, resulting in the exclusion of the penalty terms corresponding to equations \ref{subeq_qubo_b} and \ref{subeq_qubo_c}.  Similarly, in the quadratic subproblem,  all binary variables are treated as constants, resulting in the exclusion of penalty terms associated with equations \ref{subeq_qubo_d} and \ref{subeq_qubo_e}, as well as the start-up cost term \ref{eq:qubo_start-up_cost}.  In this subproblem, the absolute, time-independent power operating bounds of the units are enforced, while the remaining constraints are reformulated as described earlier.

\section{Quantum Approximate Optimization Algorithm} \label{sec:QAOA}

QAOA is a hybrid quantum-classical algorithm designed to solve combinatorial optimization problems \cite{farhi2014quantum}. It leverages the adiabatic theorem to approximate solutions for problems that are difficult to solve exactly in a classical framework. QAOA acts on an initial state, which is a known ground state of the mixer Hamiltonian $H_0$, evolving it to a final state, which is a previously unknown ground state of the problem Hamiltonian $H_P$. The two Hamiltonians must not commute, and the adiabatic process requires a slow enough evolution of the state with time $t$, resulting in the Hamiltonian
\[H(t) = (1-t) \cdot H_0 + t \cdot H_P \]
The quantum evolution is given by the unitary obtained exponentiating the Hamiltonian, \textit{i.e.}
\[e^{\frac{-i H(t)}{\hbar}} = e^{\frac{-i}{\hbar} \left[ (1-t) H_0 + t H_P \right] } \]
Since the two Hamiltonians do not commute, the Trotter-Suzuki formula is used, such that the approximation converges in the limit $p = \infty$:
\[e^{\frac{-i}{\hbar} \left[ (1-t) H_0 + t H_P \right]} \approx \left( e^{-i \frac{(1-t) H_0}{\hbar p}}  e^{-i \frac{t H_P}{\hbar p}} \right)^p\]
The two terms dependent on $t$ are replaced by the parameters $\gamma$ and $\beta$, which also absorb the $\frac{1}{\hbar p}$ term. The parameters are variationally optimized to converge to the ground state of the problem Hamiltonian. $\gamma$ and $\beta$ are vectors with length $p$, which is proportional to the runtime/depth of the QAOA circuit, corresponding to the number of iterations for the problem and mixer Hamiltonians.
\[e^{\frac{-i H(t)}{\hbar}} \approx \prod_{j=1}^p \left( e^{-i \beta_j H_0}  e^{-i \gamma_j H_P} \right) \]

\subsection{Quadratic Unconstrained Binary Optimization}

QAOA searches for approximate solutions to quadratic unconstrained binary optimization (QUBO) problems in polynomial time \cite{blekos2024review}. 
 
QUBO is a special case of Quadratic Programming, and, in its general form, for $n$ variables, is described as:
\begin{equation}
    \min_{\mathbf{x} \in \{0,1\}^n} (\mathbf{x}^T \mathbf{A} \mathbf{x} + \mathbf{b}^T \mathbf{x})
    \label{OGQ}
\end{equation}
where $b$ is a vector of linear terms in $\mathbb{R}^n$ and $A$ is a matrix of quadratic terms in $\mathbb{R}^{n\cdot n}$. 

By doing $x_i = (1-z_i)/2$ for $z_i \in \{-1,+1\}$, equation \ref{OGQ} becomes 
\begin{equation}
    \min_{\mathbf{z} \in \{-1,+1\}^n} (\mathbf{z}^T \mathbf{Q} \mathbf{z} + \mathbf{c}^T \mathbf{z})
    \label{QUBOeqmin}
\end{equation}
which represents an Ising spin glass model, with $\mathbf{Q}$ and $\mathbf{c}$ easily be computed from equation \ref{OGQ}.

Equation \ref{QUBOeqmin} can be translated into a Hamiltonian for an $n$-qubit system, replacing $z_i$ by the Pauli operator $\sigma_Z^i$ acting on the $i^{th}$ qubit, and each term of the form $z_iz_j$ by $\sigma_Z^i \otimes \sigma_Z^j$.

Given a problem's QUBO formulation, the variational form of QAOA is constructed by preparing the ground state of $H_0$, followed by $p$ iterations of the parameterized unitaries generated by $H_P$ and $H_0$.  

The unitary generated by $e^{-i \gamma_j H_p}$ is given by 
\begin{equation}
    U_P(\gamma_{j}) = 
    e^{-i \gamma_{j}(\sum_{i=1}^{n} c_i\sigma_Z^i + \sum_{i,k=1}^{n} Q_{ik} \sigma^i_Z \otimes \sigma^k_Z)}
    \label{eq:costham}
\end{equation}
The mixer Hamiltonian uses the Pauli operator $\sigma_X$, such that it does not commute with $H_P$.  The unitary generated by $e^{-i \beta_j H_0}$ is given by
\begin{equation}
    U_0(\beta_{j}) = e^{-i \beta_{j}\sum_{i=1}^{n} \sigma^i_X}
\end{equation}
The ground state of $H_0$ is the uniform superposition $|+\rangle ^{\otimes n}$, which is prepared by applying an Hadamard gate to each qubit. For a given $p$, the variational form is hence defined as:
\begin{equation}
    U(\beta, \gamma) = \left[ \prod_{j=1}^{p} U_0(\beta_j) U_P(\gamma_j)  \right] H^{\otimes n}
    \label{eq:QAOA_varform}
\end{equation}
Probability amplitudes are redistributed among basis states. States resulting in higher costs have their amplitudes transferred to states that minimize the cost.

\subsection{Warm-Start}

Warm-starting leverages the solution of a relaxed QUBO problem as the initial state for QAOA \cite{Egger_2021}.
A relaxed solution does not fully satisfy all the constraints of the original problem, rendering it simpler to solve.
In the case of QAOA, the solution to the continuous relaxation of the problem is used.

Empirical results \cite{Egger_2021} suggest that warm-starting QAOA is particularly advantageous at low depth, which is especially important in the NISQ era.

In Warm-Start QAOA, the initial state is replaced by:
\vspace{-5pt}
\begin{equation}
    | \phi^* \rangle = \bigotimes_{i=1}^{n}R_y(\theta_i) |0\rangle_i
\end{equation}
which applies rotations $R_y(\theta_i)$, \(\theta_i = 2 \text{arcsin} (\sqrt{c^*_i})\), with $c^*_i$ denoting the value of variable $i$ for the relaxed problem. The mixer Hamiltonian $H_0$ is redefined accordingly, such that it accepts $|\phi^* \rangle$ as the ground state.

It is important to note that for certain combinatorial problems, such as specific instances of MAXCUT, the standard QAOA with constant depth ($p$) does not surpass the performance of the best classical polynomial-time algorithms \cite{bravyi2020obstacles}. However, for these same MAXCUT instances, warm-start optimization matches or exceeds the performance of the best classical polynomial-time algorithms \cite{Egger_2021}.

\section{Highly Constrained UC Algorithm} \label{sec:Algorithm_Implemented}

The proposed algorithm is an iterative hybrid approach: 
\begin{description}
    \item[Classical optimizer] solves for the individual powers of each unit, $p_{i,t}$, searching in the space defined by all $N$ units and all $T$ time instants. It complies with the unit activations, given by $y_{i,t}$, which are evaluated using QAOA.

    \item[QAOA] solves for the activation of the production units for a given time instant $t$, denoted by $y_t$; solving simultaneously for all $T$ time instants would require $\mathcal{O}(NT)$ qubits, which surpasses current hardware and simulators' capabilities.

    In warm-start QAOA, where a relaxed solution of the QUBO problem is used as the initial state for QAOA, this relaxed solution is computed using a classical solver.

\end{description}

Algorithm \ref{alg:HCUC} describes the proposed approach in pseudo-code, and figure \ref{fig:diagram2} displays the flowchart.

\begin{algorithm*}[hbt!]
\caption{Highly Constrained UC} \label{alg:HCUC}
\tcc{
$p$ is a $N * T$ vector, holding all $p_{i,t}$ \\
$p_{i,.}$ refers to all $p_{i,1} \ldots p_{i,T}$, for a given unit $i$ \\
$p_{.,t}$ refers to all $p_{1,t} \ldots p_{N,t}$, for a given time instant $t$ \\
$y$ is a $N * T$ vector, holding all $y_{i,t}$
}
\BlankLine
\tcp{Initial guess for the power output of each unit}
\lForAll{$i \in \{1, \ldots, n\}$}{$p_{i,.} \gets p^{max}_i$} \label{line:init_p}
\BlankLine
\label{line:rel_min}
\tcp{Optimize for initial unit activation}
\lForAll{$t \in \{1, \ldots, T\}$}{
$y_{.,t} \gets \text{QAOA} (t, p_{.,t})$ 
} \label{line:rel_qaoa}
\BlankLine
\tcp{Iterate to converge the unit activation solution}
\For{$N_{it}$ iterations}{ \label{line:Nit_for}
$p \gets \text{Minimize} (p,y)$ \;
\lForAll{$t \in \{1, \ldots, T\}$}{
$y_{.,t} \gets \text{QAOA} (t, p_{.,t})$ }
} \label{line:Nit_end}
\BlankLine
\tcp{Final power  optimization }
$p \gets \text{FinalMinimize} (p,y)$ \;
 \label{line:final_p}

\end{algorithm*}

\tikzstyle{startstop} = [rectangle, rounded corners, 
minimum width=2.5cm, 
minimum height=0.9cm,
align=center,
text centered, 
draw=black, 
fill=red!20]

\tikzstyle{process} = [rectangle, 
minimum width=0.9cm, 
minimum height=0.9cm, 
text centered, 
text width=2cm, align=center,
draw=black, 
fill=blue!20]

\tikzstyle{pr} = [rectangle, 
minimum width=0.9cm, 
minimum height=0.9cm, 
text centered, 
text width=1.5cm, align=center,
draw=black, 
fill=blue!20]

\tikzstyle{decision} = [diamond, 
minimum width=2.5cm, 
minimum height=2.5cm,  align=center, 
text centered, 
draw=black, 
fill=green!20]
\tikzstyle{arrow} = [thick,->,>=stealth]

\begin{figure}[hbt!]
\centering
\begin{tikzpicture}[node distance=1.1cm]

    \node (start) [startstop] {\footnotesize Start};
    \node (pro1) [process, below of=start, yshift=-0.6cm] {\footnotesize  $p_{i,t}$ = $p_{i}^{\max}$; \\$t=0$; \\ $n_{it}=0$;};
    \node (dec0) [decision, below of=pro1, yshift=-1.5cm] { \scriptsize $t \leq T?$};
    \node (qaoa1) [pr, left of=dec0, xshift=-1.45cm, yshift=1.53cm] { \footnotesize QAOA; \\ $t  \mathrel{+}= 1$;};
    \node (dec1) [decision, below of=dec0, yshift=-2cm] {\footnotesize $n_{it} \leq N_{it}?$};
    \node (op_p) [process, right of=dec1, xshift=1.9cm] {\footnotesize Optimize $p$; \\ $n_{it}  \mathrel{+}= 1$; \\ $t=0$;};
    \node (op_p_f) [process, below of=dec1, yshift=-1.3cm] {\footnotesize Final $p$ \\ optimization};
    \node (stop) [startstop, below of=op_p_f, yshift=-0.5cm] {\footnotesize End};

    \draw [arrow] (start) -- (pro1);
    \draw [arrow] (pro1) -- (dec0);
    \draw [arrow] (dec0) -| node[anchor=north] {\scriptsize yes} (qaoa1);
    \draw [arrow] (dec0) -- node[anchor=east] {\scriptsize no} (dec1);
    \draw [arrow] (qaoa1) -| (dec0);
    \draw [arrow] (dec1) -- node[anchor=east] {\scriptsize no} (op_p_f);
    \draw [arrow] (dec1) -- node[anchor=south] {\scriptsize yes} (op_p);
    \draw [arrow] (op_p) |- (dec0);
    \draw [arrow] (op_p_f) -- (stop);

\end{tikzpicture}
\caption{ Flowchart of the hybrid UC algorithm}
\label{fig:diagram2}
\end{figure}
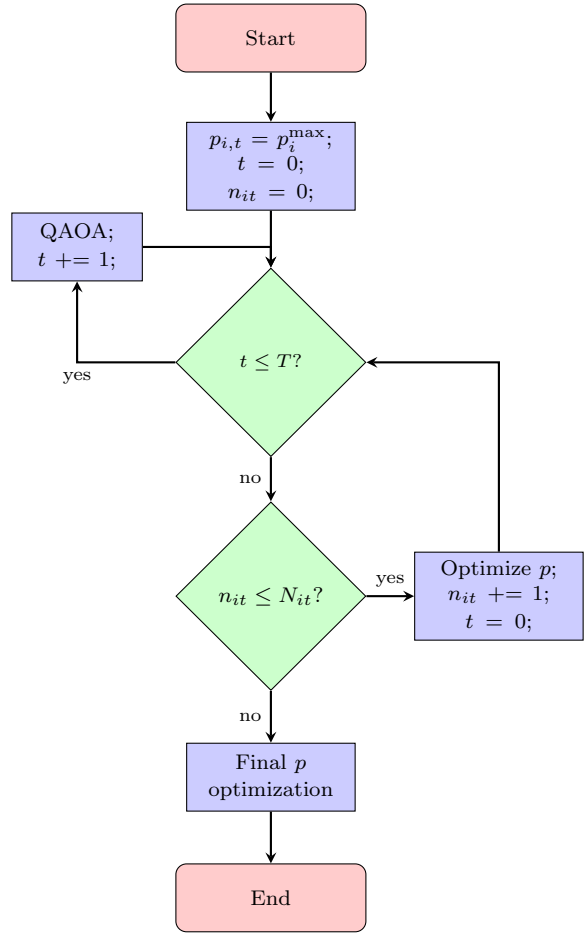

The first step entails establishing a good initial estimate for the scheduling of power generating units (lines \ref{line:init_p} and \ref{line:rel_qaoa}). Each individual power unit is set to its maximum power capacity. Subsequently, QAOA algorithm is utilized to identify the combination of units that minimize the objective function based on the power levels established in the previous step. As previously noted, QAOA handles each time instant separately due to limitations on both current quantum hardware and quantum circuits' classical simulators.

Lines \ref{line:Nit_for} to \ref{line:Nit_end} implement an iterative loop designed to determine the activation state of each unit based on an estimated power level distribution. The term `estimate' is employed here given that the objective function used to calculate the power levels yields low values for the penalty terms $\lambda$ (sub-section \ref{subsec:QUBO_formulation}). Within this loop, the power assigned to each unit at every time step is evaluated classically using the current operation schedule, $y$. Subsequently, QAOA optimizes the operation schedule to produce an updated distribution of power outputs, $p$. It should be noted that this approach of starting with low values for the penalty terms $\lambda$  was chosen based on the quality of its results.

The power solution obtained at the end of the loop does not fully satisfy the constraints. To address this issue the final units' power distribution is classically optimized using the current operation schedule $y$ and larger penalty terms $\lambda$ --  line \ref{line:final_p}.


\section{Experimental Methodology} \label{sec:Methodology}

\subsection{UC instances}

To assess the algorithm's efficacy in approximating solutions to the UC problem, a total of six different problems were examined, with four, ten, and twelve power generation units. Three time steps ($T=3$) were used for all problems.
These problems are detailed in the appendix, tables \ref{tabela:prob6} to \ref{tabela:prob11}.

For four-unit problems, the parameter $N_{it}$ in Algorithm \ref{alg:HCUC} was set to 3. For the ten-unit problems, it was set to 6. In UC\_12a $N_{it}$  was set to 12. For UC\_12b, it was set to 7 when using standard QAOA and 6 when using the warm-start optimization. The values for the twelve-unit instances were determined by limitations in computational resources.

\subsection{Experimental configuration}

All experiments were performed by resorting to classical simulation of the quantum circuits using Qiskit 0.45.0  \cite{Qiskit}.  For the classical component of the proposed algorithm, the minimize function from SciPy 1.11.4 \cite{virtanen2020scipy} was used.

Table \ref{table1} presents detailed system information.

\begin{table}[H]
\centering
\renewcommand{\arraystretch}{1.2} 
\setlength{\tabcolsep}{10pt}     
\begin{tabular}{ |c|c| } 
\hline
\hline
Python version & 3.10.8\\
\hline
Python compiler & GCC 11.3.0 \\
\hline
Python build & main Apr 24 2023 17:09:49  \\
\hline
OS & Linux \#1 SMP Tue Feb 21  \\ 
& 04:20:52 EST 2023 \\ 
\hline
CPU & x86\_64 \\
\hline
Memory (Gb) & 503.24 GB \\
\hline
\hline
\end{tabular}
\caption{System information}
\label{table1}
\end{table}

\subsection{Hybrid algorithm setup} \label{subsec:algorithm_setup}

Qiskit's implementation of QAOA was used for all experiments. This function handles the conversion from a QUBO to the respective Hamiltonian.  It also supports warm starting, using the CplexOptimizer \cite{cplex2009v12} to compute a solution for the relaxed QUBO problem.

Each power generation unit state (off/on) is represented by one qubit in the QAOA circuit, requiring $N$ qubits. Qiskit's QAOA function uses nine additional qubits to enforce all constraints, including the spinning reserve. The total number of qubits is thus $N+9$.

 COBYLA classical solver is employed to variationally optimize QAOA's parameters, $\gamma$ and $\beta$. 
 The  \texttt{maxiter}  parameter, which specifies the maximum number of iterations allowed for the optimizer, is set to its default value (within Qiskit) of 1000. 

For the quantum optimization, constraints' weights ($\lambda$ -- sub-section \ref{subsec:QUBO_formulation}) for minimum up and downtime in the QUBO formulation are set to 100, while those for load and spinning reserve constraints are assigned a weight of 1.  

Given the limitations of computational resources,  a single layer was used for the trotterization, \textit{i.e.} parameter $p$ in equation \ref{eq:QAOA_varform} was set to 1.

For the classical minimizations, the COBYLA optimizer was selected as the minimizer.  Here, the  $\lambda$ terms define two objective functions: within the $N_{it}$ loop of the algorithm these are set to $0.5$. However, at the end of this loop, the $\lambda$ terms are increased to  $10^4$ to compute the final power distribution.  Additionally, the \texttt{maxiter} parameter is set to  $10^4$ during the $N_{it}$  loop and to $10^7$ in the latter case.

\subsection{Reference classical solver}

The performance of the proposed algorithm was evaluated by benchmarking it against a full classical solver: Gurobi Optimizer version 10.0.2 build v10.0.2rc0 (win64) \cite{gurobi}. 

The best classical (approximate) solutions, used as references for assessing the proposed hybrid approach for the evaluated UC instances, are presented in table \ref{tabela:best_class_10}.

\section{Results} \label{sec:Results}

\subsection{Objective function}

The detailed solutions obtained for the six UC instances using both standard and warm-start QAOA algorithms are presented in the appendix, tables \ref{tabela:results_std} and \ref{tabela:results_ws}, respectively. It is clear from this data that all constraints are satisfied. 

\begin{figure} [H] 
    \centering
    \includegraphics[width=7.8cm]{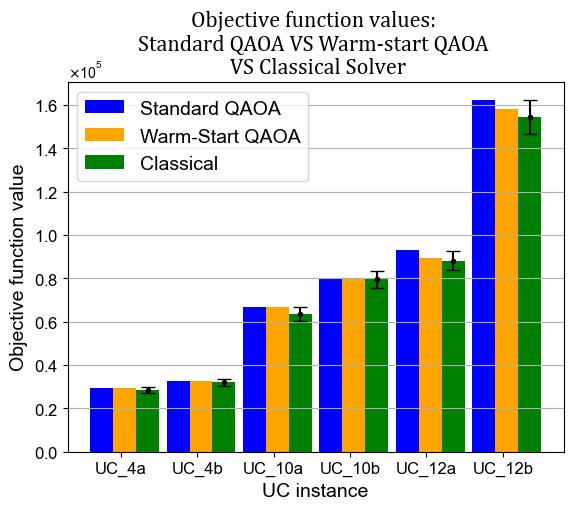}
\caption{Objective values for the six UC instances : Standard QAOA vs Warm-start QAOA vs Reference classical solver. The error bar represents a 5.1 percent relative error compared to the classical solution.}
\label{fig:results}
\end{figure}

Table \ref{tabela:results_comp} presents the objective function values obtained with the standard QAOA, warm-start QAOA, and the reference classical solver. Additionally, these results are graphically represented in figure \ref{fig:results}.

\begin{table}[H]
\centering
\renewcommand{\arraystretch}{1.2} 
\setlength{\tabcolsep}{10pt}     
\begin{tabular}{|c|d|d|d|}
\hline
\hline
 Instance &   \multicolumn{1}{c|}{\textrm{QAOA}}   &  \multicolumn{1}{c|}{\textrm{WS-QAOA}}   &  \multicolumn{1}{c|}{\textrm{Classical}}   \\ 
\hline
$\text{UC\_4a}$ & 29.3 & 29.3 & 28.3 \\ \hline
$\text{UC\_4b}$ & 32.4 & 32.4 & 32.0 \\ \hline
$\text{UC\_10a}$ & 66.8 & 66.8 & 63.5 \\ \hline
$\text{UC\_10b}$ & 80.2 & 80.2 & 79.6 \\ \hline
$\text{UC\_12a}$ & 93.1& 89.3 & 88.0 \\ \hline
$\text{UC\_12b}$ & 162.6 & 158.1 & 154.5 \\ \hline
\hline
\end{tabular}
\caption{Objective values (in thousands) for the six UC instances: Standard QAOA vs Warm-start QAOA vs Reference classical solver.}
\label{tabela:results_comp}
\end{table}

A comparison of the QAOA and classical algorithms' performance shows that the hybrid algorithm yields a higher objective function cost across all six UC instances compared to the classical approach. However, hybrid warm-start solutions remain within a 5.1\% margin of the classical solution.

Advantages of warm-start QAOA, with respect to standard QAOA, only surface for the larger UC instances, with 12 power units.  For instances UC\_12a and UC\_12b warm-start yields a smaller cost than the standard approach.  For all the remaining instances, the solutions of both algorithms are the same.

Table \ref{tabela:it_coverge_comp} summarizes the number of iterations (of the $N_{it}$ loop) required for the QAOA solution to converge for each problem, comparing the standard QAOA and the warm-start QAOA. It is important to note that convergence, in this context, is defined as the point at which the QAOA consistently produces the same solution from that iteration onward. Iteration $0$ represents the QAOA solution corresponding to the initial guess for the power output of each unit.

\begin{table}[H]
\centering
\renewcommand{\arraystretch}{1.2} 
\setlength{\tabcolsep}{10pt}     
\begin{tabular}{|c|d|d|}
\hline
\hline
 Instance &   \multicolumn{1}{c|}{\textrm{QAOA}}   &  \multicolumn{1}{c|}{\textrm{WS-QAOA}}      \\ 
\hline
$\text{UC\_4a}$ & 1 & 1 \\ \hline
$\text{UC\_4b}$ & 0 & 0 \\ \hline
$\text{UC\_10a}$ & 3 & 3 \\ \hline
$\text{UC\_10b}$ & 1 & 1  \\ \hline
$\text{UC\_12a}$ & - & 0  \\ \hline
$\text{UC\_12b}$ & - & 1  \\ \hline
\hline
\end{tabular}
\caption{Iteration (of the $N_{it}$ loop) at which the QAOA solution converged: Standard QAOA vs. Warm-Start QAOA}
\label{tabela:it_coverge_comp}
\end{table}

  Once more, the advantages of the warm-start optimization are evident in the 12-unit instances. In this instances, under standard QAOA, the solution failed to converge, whereas, with the warm-start approach, convergence was achieved rapidly.

  Here, the focus is solely on the convergence of the QAOA solution, as the purpose of the  $N_{it}$  loop is to determine the optimal unit activation state.

\subsection{Time complexity remark}

In the classical component of the hybrid algorithm, the COBYLA optimizer was employed to minimize the objective function. The time complexity of COBYLA is $\mathcal{O}(m^2)$, where $m$ denotes the dimension of the vector being optimized. Here, $m$ corresponds to $NT$. This complexity assumes a fixed maximum number of iterations the optimizer is allowed to go through \cite{liu2024quantum}, which is described in sub-section \ref{subsec:algorithm_setup}. Consequently, the classical part of the hybrid algorithm implemented has a time complexity of $\mathcal{O}((NT)^2)$.

As previously mentioned, the standard QAOA algorithm has a polynomial time complexity \cite{blekos2024review}. 
The QAOA warm-start optimization using the Cplex optimizer to solve the relaxed QUBO problem, also runs in polynomial time \cite{Egger_2021}. Note that the COBYLA optimizer was also used here as the classical minimizer. In the QAOA context, the dimension of the vector being optimized corresponds to $m=2p$, where $p$ represents the length  of vectors $\gamma$ and $\beta$.  As a result, the time complexity of the optimization process is $\mathcal{O}(p^2)$. Given that the QAOA algorithm requires $N+9$ qubits, the time complexity of the quantum part of the hybrid algorithm implemented is $\mathcal{O}(\text{poly}(p  N))$.


Therefore, the overall time complexity of the algorithm is $\mathcal{O}(\text{poly}(p  N) + (NT)^3)$.

The classical solver used for the reference solutions \cite{gurobi} uses a branch and bound approach whose worst-case complexity is known to be exponential \cite{zhang1996branch, thakoor2009computation, denat2024}. Although an average time complexity characterization for this problem could not be found, the solver is known to exhibit a sub-exponential, polynomial, average-case time complexity for related problems \cite{borst2023integrality, denat2024}. 

\section{Conclusion}

This article devised a hybrid quantum-classical approach to solving the UC problem in polynomial time. The algorithm combines classical optimization methods for calculating individual unit power outputs with the QAOA to select the optimal combination of units. A set of constraints is considered, including the spinning reserve constraint, which has not previously been addressed in hybrid quantum-classical environments.  The results show that the solutions achieved using a single-layer warm-start QAOA ($p=1$)  are within a 5.1\% margin of its classical counterpart, with warm-start optimization proving particularly beneficial for larger problem instances. In these problem instances, not only was higher accuracy achieved, but the QAOA solution also converged in fewer iterations.

Due to computational resource limitations, the behavior of the hybrid algorithm could not be investigated for more than one layer in QAOA. Nevertheless, it is expected that the algorithm's performance improves as $p$ increases.

Future work should involve testing the algorithm on real quantum machines and investigating the potential for adaptive penalty adjustments in the quadratic unconstrained problem. Specifically, penalties could start lower in the initial iterations and progressively increase in later stages, potentially enhancing convergence and solution accuracy. Moreover, the potential to replace the global parameters $\gamma$ and $\beta$ with individual parameters for each gate in the quantum circuit should be studied. This approach could enhance the circuit's expressiveness without necessitating an increase in the number of layers.

\begin{acknowledgments}
This work is financed by National Funds through the Portuguese funding agency, FCT - Fundação para a Ciência e a Tecnologia, within projects \href{https://doi.org/10.54499/LA/P/0063/2020}{LA/P/0063/2020}, IBEX with reference 10.54499/PTDC/CCI-COM/4280/2021 and grants UI/BD/152698/2022 and 11313/BI-L-EM\_B3/2024. Moreover, project 2023.09537.CPCA.A1 provided the required computing resources.
\end{acknowledgments}

\newpage
\bibliographystyle{unstrturl}
\bibliographystyle{IEEEtran}
\bibliography{main_arXiv}

\providecommand{\noopsort}[1]{}\providecommand{\singleletter}[1]{#1}%
\begin{thebibliography}{10}
\providecommand{\url}[1]{#1}
\csname url@samestyle\endcsname
\providecommand{\newblock}{\relax}
\providecommand{\bibinfo}[2]{#2}
\providecommand{\BIBentrySTDinterwordspacing}{\spaceskip=0pt\relax}
\providecommand{\BIBentryALTinterwordstretchfactor}{4}
\providecommand{\BIBentryALTinterwordspacing}{\spaceskip=\fontdimen2\font plus
\BIBentryALTinterwordstretchfactor\fontdimen3\font minus \fontdimen4\font\relax}
\providecommand{\BIBforeignlanguage}[2]{{%
\expandafter\ifx\csname l@#1\endcsname\relax
\typeout{** WARNING: IEEEtran.bst: No hyphenation pattern has been}%
\typeout{** loaded for the language `#1'. Using the pattern for}%
\typeout{** the default language instead.}%
\else
\language=\csname l@#1\endcsname
\fi
#2}}
\providecommand{\BIBdecl}{\relax}
\BIBdecl

\bibitem{padhy2004unit}
\BIBentryALTinterwordspacing
N.~P. Padhy, ``Unit commitment-a bibliographical survey,'' \emph{IEEE Transactions on power systems}, vol.~19, no.~2, pp. 1196--1205, 2004. [Online]. Available: \url{http://ieeexplore.ieee.org/document/1295033/}
\BIBentrySTDinterwordspacing

\bibitem{ajagekar2019quantum}
\BIBentryALTinterwordspacing
A.~Ajagekar and F.~You, ``Quantum computing for energy systems optimization: Challenges and opportunities,'' \emph{Energy}, vol. 179, pp. 76--89, 2019. [Online]. Available: \url{https://linkinghub.elsevier.com/retrieve/pii/S0360544219308254}
\BIBentrySTDinterwordspacing

\bibitem{datta2013unit}
\BIBentryALTinterwordspacing
D.~Datta, ``Unit commitment problem with ramp rate constraint using a binary-real-coded genetic algorithm,'' \emph{Applied Soft Computing}, vol.~13, no.~9, pp. 3873--3883, 2013. [Online]. Available: \url{https://linkinghub.elsevier.com/retrieve/pii/S1568494613001579}
\BIBentrySTDinterwordspacing

\bibitem{tseng1996power}
C.-L. Tseng, \emph{On power system generation unit commitment problems}.\hskip 1em plus 0.5em minus 0.4em\relax University of California, Berkeley, 1996.

\bibitem{1270468}
\BIBentryALTinterwordspacing
X.~Guan, Q.~Zhai, and A.~Papalexopoulos, ``Optimization based methods for unit commitment: Lagrangian relaxation versus general mixed integer programming,'' in \emph{2003 IEEE Power Engineering Society General Meeting (IEEE Cat. No.03CH37491)}, vol.~2, 2003, pp. 1095--1100 Vol. 2. [Online]. Available: \url{http://ieeexplore.ieee.org/document/1270468/}
\BIBentrySTDinterwordspacing

\bibitem{farhi2014quantum}
\BIBentryALTinterwordspacing
E.~Farhi, J.~Goldstone, and S.~Gutmann, ``A quantum approximate optimization algorithm,'' 2014. [Online]. Available: \url{http://arxiv.org/abs/1411.4028}
\BIBentrySTDinterwordspacing

\bibitem{blekos2024review}
\BIBentryALTinterwordspacing
K.~Blekos, D.~Brand, A.~Ceschini, C.-H. Chou, R.-H. Li, K.~Pandya, and A.~Summer, ``A review on quantum approximate optimization algorithm and its variants,'' \emph{Physics Reports}, vol. 1068, pp. 1--66, 2024. [Online]. Available: \url{https://linkinghub.elsevier.com/retrieve/pii/S0370157324001078}
\BIBentrySTDinterwordspacing

\bibitem{Egger_2021}
\BIBentryALTinterwordspacing
D.~J. Egger, J.~Mareček, and S.~Woerner, ``Warm-starting quantum optimization,'' \emph{Quantum}, vol.~5, p. 479, 2021. [Online]. Available: \url{https://quantum-journal.org/papers/q-2021-06-17-479/}
\BIBentrySTDinterwordspacing

\bibitem{koretsky2021adapting}
\BIBentryALTinterwordspacing
S.~Koretsky, P.~Gokhale, J.~M. Baker, J.~Viszlai, H.~Zheng, N.~Gurung, R.~Burg, E.~A. Paaso, A.~Khodaei, R.~Eskandarpour \emph{et~al.}, ``Adapting quantum approximation optimization algorithm (qaoa) for unit commitment,'' in \emph{2021 IEEE International Conference on Quantum Computing and Engineering (QCE)}.\hskip 1em plus 0.5em minus 0.4em\relax IEEE, 2021, pp. 181--187. [Online]. Available: \url{https://ieeexplore.ieee.org/document/9605315/}
\BIBentrySTDinterwordspacing

\bibitem{nikmehr2022quantum}
\BIBentryALTinterwordspacing
N.~Nikmehr, P.~Zhang, and M.~A. Bragin, ``Quantum distributed unit commitment: An application in microgrids,'' \emph{IEEE transactions on power systems}, vol.~37, no.~5, pp. 3592--3603, 2022. [Online]. Available: \url{https://ieeexplore.ieee.org/document/9677977/}
\BIBentrySTDinterwordspacing

\bibitem{mahroo2022hybrid}
\BIBentryALTinterwordspacing
R.~Mahroo and A.~Kargarian, ``Hybrid quantum-classical unit commitment,'' in \emph{2022 IEEE Texas Power and Energy Conference (TPEC)}.\hskip 1em plus 0.5em minus 0.4em\relax IEEE, 2022, pp. 1--5. [Online]. Available: \url{https://ieeexplore.ieee.org/document/9750763/}
\BIBentrySTDinterwordspacing

\bibitem{feng2022novel}
\BIBentryALTinterwordspacing
F.~Feng, P.~Zhang, M.~A. Bragin, and Y.~Zhou, ``Novel resolution of unit commitment problems through quantum surrogate lagrangian relaxation,'' \emph{IEEE Transactions on Power Systems}, vol.~38, no.~3, pp. 2460--2471, 2022. [Online]. Available: \url{https://ieeexplore.ieee.org/document/9793720/}
\BIBentrySTDinterwordspacing

\bibitem{stein2023combining}
\BIBentryALTinterwordspacing
J.~Stein, J.~Jojo, A.~Farea, D.~Bucher, P.~Altmann, and C.~Linnhoff-Popien, ``Combining the qaoa and hhl algorithm to achieve a substantial quantum speedup for the unit commitment problem,'' 2023. [Online]. Available: \url{http://arxiv.org/abs/2305.08482}
\BIBentrySTDinterwordspacing

\bibitem{bravyi2020obstacles}
\BIBentryALTinterwordspacing
S.~Bravyi, A.~Kliesch, R.~Koenig, and E.~Tang, ``Obstacles to variational quantum optimization from symmetry protection,'' \emph{Physical review letters}, vol. 125, no.~26, p. 260505, 2020. [Online]. Available: \url{https://link.aps.org/doi/10.1103/PhysRevLett.125.260505}
\BIBentrySTDinterwordspacing

\bibitem{Qiskit}
\BIBentryALTinterwordspacing
A.~Javadi-Abhari, M.~Treinish, K.~Krsulich, C.~J. Wood, J.~Lishman, J.~Gacon, S.~Martiel, P.~D. Nation, L.~S. Bishop, A.~W. Cross, B.~R. Johnson, and J.~M. Gambetta, ``Quantum computing with {Q}iskit,'' 2024. [Online]. Available: \url{https://arxiv.org/abs/2405.08810}
\BIBentrySTDinterwordspacing

\bibitem{virtanen2020scipy}
\BIBentryALTinterwordspacing
P.~Virtanen, R.~Gommers, T.~E. Oliphant, M.~Haberland, T.~Reddy, D.~Cournapeau, E.~Burovski, P.~Peterson, W.~Weckesser, J.~Bright \emph{et~al.}, ``Scipy 1.0: fundamental algorithms for scientific computing in python,'' \emph{Nature methods}, vol.~17, no.~3, pp. 261--272, 2020. [Online]. Available: \url{https://www.nature.com/articles/s41592-019-0686-2}
\BIBentrySTDinterwordspacing

\bibitem{cplex2009v12}
I.~I. Cplex \emph{et~al.}, \emph{V12. 1: User’s Manual for CPLEX}.\hskip 1em plus 0.5em minus 0.4em\relax Armonk, NY, USA, 2009, vol.~46, no.~53.

\bibitem{gurobi}
\BIBentryALTinterwordspacing
{Gurobi Optimization, LLC}, ``{Gurobi Optimizer Reference Manual},'' 2024. [Online]. Available: \url{https://www.gurobi.com}
\BIBentrySTDinterwordspacing

\bibitem{liu2024quantum}
\BIBentryALTinterwordspacing
Y.~Liu, J.~Liu, J.~R. Raney, and P.~Wang, ``Quantum computing for solid mechanics and structural engineering--a demonstration with variational quantum eigensolver,'' \emph{Extreme Mechanics Letters}, vol.~67, p. 102117, 2024. [Online]. Available: \url{https://linkinghub.elsevier.com/retrieve/pii/S2352431623001633}
\BIBentrySTDinterwordspacing

\bibitem{zhang1996branch}
\BIBentryALTinterwordspacing
W.~Zhang, \emph{Branch-and-bound search algorithms and their computational complexity}.\hskip 1em plus 0.5em minus 0.4em\relax University of Southern California, Information Sciences Institute, 1996. [Online]. Available: \url{https://apps.dtic.mil/sti/citations/ADA314598}
\BIBentrySTDinterwordspacing

\bibitem{thakoor2009computation}
\BIBentryALTinterwordspacing
N.~Thakoor, V.~Devarajan, and J.~Gao, ``Computation complexity of branch-and-bound model selection,'' in \emph{2009 IEEE 12th International Conference on Computer Vision}.\hskip 1em plus 0.5em minus 0.4em\relax IEEE, 2009, pp. 1895--1900. [Online]. Available: \url{http://ieeexplore.ieee.org/document/5459420/}
\BIBentrySTDinterwordspacing

\bibitem{denat2024}
\BIBentryALTinterwordspacing
T.~Denat, A.~Harutyunyan, N.~Melissinos, and V.~Paschos, ``{Average-case complexity of a branch-and-bound algorithm for Min Dominating Set},'' \emph{Discrete Applied Mathematics}, vol. 345, pp. 4--8, 2024. [Online]. Available: \url{https://linkinghub.elsevier.com/retrieve/pii/S0166218X23004353}
\BIBentrySTDinterwordspacing

\bibitem{borst2023integrality}
\BIBentryALTinterwordspacing
S.~Borst, D.~Dadush, and D.~Mikulincer, ``Integrality gaps for random integer programs via discrepancy,'' in \emph{Proceedings of the 2023 Annual ACM-SIAM Symposium on Discrete Algorithms (SODA)}.\hskip 1em plus 0.5em minus 0.4em\relax SIAM, 2023, pp. 1692--1733. [Online]. Available: \url{https://epubs.siam.org/doi/book/10.1137/1.9781611977554}
\BIBentrySTDinterwordspacing

\end{thebibliography}


\appendix

\begin{figure*} [h]

\vspace{-20pt}

\newpage

\section{ UC instances}
\end{figure*}

\begin{center}
\begin{table*}[!htbp]
\centering
\renewcommand{\arraystretch}{1.1} 
\setlength{\tabcolsep}{10pt}     
\begin{tabular}{|c|c|c|c|c|}
\hline \hline
i & 1 & 2 & 3 & 4 \\ \hline
$p^{\min}_{i}$ & 10 & 10 & 25 & 150 \\
$p^{\max}_{i}$ & 55 & 100 & 85 & 500 \\
$A_i$ & 670 & 450 & 735 & 370 \\
$B_i$ & 25.92 & 16.5 & 16.7 & 22.26 \\
$C_i$ & 0.00413 & 0.002 & 0.00398 & 0.00712 \\
$D_i$ & 10 & 15 & 12 & 9 \\
$E_i$ & 12 & 18 & 13 & 14 \\
$F_i$ & 1 & 2 & 2 & 1 \\
$T^{down}_{i}$ & 4 & 1 & 1 & 3 \\
$T^{up}_{i}$ & 1 & 1 & 2 & 23 \\
$r^{dn}_{i}$ & 10 & 40 & 20 & 100 \\
$r^{up}_{i}$ & 5 & 30 & 15 & 90 \\ \hline
\multicolumn{2}{|c|}{t} & 1& 2& 3 \\ \hline
\multicolumn{2}{|c|}{$L_t$} & 350 & 300 & 500  \\
\multicolumn{2}{|c|}{spinning reserve} & 20 &10&30\\ \hline \hline
\end{tabular}
\caption{Instance UC\_4a}
\label{tabela:prob6}
\end{table*}
\end{center}

\vspace{-80pt}

\begin{center}
\begin{table*}[htbp]
\centering
\renewcommand{\arraystretch}{1.1} 
\setlength{\tabcolsep}{10pt}     
\begin{tabular}{|c|c|c|c|c|}
\hline \hline
i & 1 & 2 & 3 & 4 \\ \hline
$p^{\min}_{i}$ & 150 & 20 & 25 & 20 \\
$p^{\max}_{i}$ & 455 & 130 & 165 & 80\\
$A_i$ & 1000 & 700 & 450 & 370 \\
$B_i$ & 16.19 & 16.5 & 16.7 & 22.26 \\
$C_i$ & 0.00048 & 0.002 & 0.00398 & 0.00712 \\
$D_i$ & 9 & 12 & 15 & 10\\
$E_i$ & 14 & 13 & 18 & 12 \\
$F_i$ & 2 & 1 & 1 & 2 \\
$T^{down}_{i}$ & 2 & 1 & 1 & 4 \\
$T^{up}_{i}$ & 3 & 2 & 1 & 1 \\
$r^{dn}_{i}$ & 100 & 30 & 40 & 10 \\
$r^{up}_{i}$ & 80 & 15 & 30 & 5 \\ \hline
\multicolumn{2}{|c|}{t} & 1& 2& 3 \\ \hline
\multicolumn{2}{|c|}{$L_t$} & 650 & 530 & 450  \\
\multicolumn{2}{|c|}{spinning reserve} & 50 & 25 & 15\\ \hline \hline
\end{tabular}
\caption{Instance UC\_4b}
\label{tabela:prob7}
\end{table*}
\end{center}

\begin{center}
\begin{table*}[htbp]
\centering
\renewcommand{\arraystretch}{1.1} 
\setlength{\tabcolsep}{5pt}     
\begin{tabular}{|c|c|c|c|c|c|c|c|c|c|c|}
\hline \hline
i & 1 & 2 & 3 & 4 & 5 & 6 & 7 & 8 & 9 & 10 \\ \hline
$p^{\min}_{i}$ & 10 & 10 & 20 & 20 & 25 & 150 & 25 & 10 & 150 & 20 \\
$p^{\max}_{i}$ & 55 & 55 & 130 & 130 & 165 & 455 & 85 & 55 & 455 & 80\\
$A_i$ & 660 & 670 & 700 & 680 & 450 & 970 & 480 & 665 & 1000 & 370 \\
$B_i$ & 25.92 & 27.76 & 16.6 & 16.5 & 19.7 & 17.26 & 27.74 & 27.27 & 16.19 & 22.26 \\
$C_i$ & 0.00413 & 0.00173 & 0.002 & 0.00211 & 0.00398 & 0.00031 & 0.0079 & 0.00222 & 0.00048 & 0.00712 \\
$D_i$ & 13 & 8 & 12 & 16 & 10 & 12 & 17 & 12 & 7 & 15\\
$E_i$ & 15 & 11 & 13 & 20 & 12 & 15 & 20 & 14 & 12 & 18 \\
$F_i$ & 1 & 2 & 3 & 1 & 2 & 2 & 1 & 2 & 1 & 3 \\
$T^{down}_{i}$ & 2 & 4 & 1 & 3 & 1 & 2 & 1 & 2 & 1 & 3 \\
$T^{up}_{i}$ & 1 & 3 & 2 & 1 & 1 & 2 & 1 & 4 & 1 & 1\\
$r^{dn}_{i}$ & 25 & 10 & 30 & 50 & 35 & 60 & 70 & 100 & 80 & 40 \\
$r^{up}_{i}$ & 80 & 20 & 20 & 40 & 35 & 50 & 15 & 80 & 50 & 30\\ \hline
\multicolumn{2}{|c|}{t} & \multicolumn{3}{|c|}{1} & \multicolumn{3}{|c|}{2} & \multicolumn{3}{|c|}{3} \\ \hline
\multicolumn{2}{|c|}{$L_t$} & \multicolumn{3}{|c|}{900} & \multicolumn{3}{|c|}{1000} & \multicolumn{3}{|c|}{1300} \\
\multicolumn{2}{|c|}{spinning reserve} & \multicolumn{3}{|c|}{20} & \multicolumn{3}{|c|}{10} & \multicolumn{3}{|c|}{30}\\ \hline \hline
\end{tabular}
\caption{Instance UC\_10a}
\label{tabela:prob8}
\end{table*}
\end{center}

\begin{center}
\begin{table*}[htbp]
\centering
\renewcommand{\arraystretch}{1.2} 
\setlength{\tabcolsep}{5pt}     
\begin{tabular}{|c|c|c|c|c|c|c|c|c|c|c|}
\hline \hline
i & 1 & 2 & 3 & 4 & 5 & 6 & 7 & 8 & 9 & 10 \\ \hline
$p^{\min}_{i}$ & 10 & 10 & 20 & 20 & 25 & 150 & 25 & 10 & 150 & 20 \\
$p^{\max}_{i}$ & 55 & 55 & 130 & 130 & 165 & 455 & 85 & 55 & 455 & 80\\
$A_i$ & 660 & 670 & 700 & 680 & 450 & 970 & 480 & 665 & 1000 & 370 \\
$B_i$ & 25.92 & 27.76 & 16.6 & 16.5 & 19.7 & 17.26 & 27.74 & 27.27 & 16.19 & 22.26 \\
$C_i$ & 0.00413 & 0.00173 & 0.002 & 0.00211 & 0.00398 & 0.00031 & 0.0079 & 0.00222 & 0.00048 & 0.00712 \\
$D_i$ & 13 & 8 & 12 & 16 & 10 & 12 & 17 & 12 & 7 & 15\\
$E_i$ & 15 & 11 & 13 & 20 & 12 & 15 & 20 & 14 & 12 & 18 \\
$F_i$ & 1 & 2 & 3 & 1 & 2 & 2 & 1 & 2 & 1 & 3 \\
$T^{down}_{i}$ & 2 & 4 & 1 & 3 & 1 & 2 & 1 & 2 & 1 & 3 \\
$T^{up}_{i}$ & 1 & 3 & 2 & 1 & 1 & 2 & 1 & 4 & 1 & 1\\
$r^{dn}_{i}$ & 25 & 10 & 30 & 50 & 35 & 60 & 70 & 100 & 80 & 40\\
$r^{up}_{i}$ & 80 & 20 & 20 & 40 & 35 & 50 & 15 & 80 & 50 & 30\\ \hline
\multicolumn{2}{|c|}{t} & \multicolumn{3}{|c|}{1} & \multicolumn{3}{|c|}{2} & \multicolumn{3}{|c|}{3} \\ \hline
\multicolumn{2}{|c|}{$L_t$} & \multicolumn{3}{|c|}{1300} & \multicolumn{3}{|c|}{1400} & \multicolumn{3}{|c|}{1200} \\
\multicolumn{2}{|c|}{spinning reserve} & \multicolumn{3}{|c|}{20} & \multicolumn{3}{|c|}{10} & \multicolumn{3}{|c|}{30}\\ \hline \hline
\end{tabular}
\caption{ Instance UC\_10b}
\label{tabela:prob9}
\end{table*}
\end{center}

\begin{center}
\begin{table*}[htbp]
\centering
\renewcommand{\arraystretch}{1.2} 
\setlength{\tabcolsep}{10pt}     
\begin{tabular}{|c|c|c|c|c|c|c|}
\hline\ hline
i & 1 & 2 & 3 & 4 & 5 & 6  \\ \hline 
$p^{\min}_{i}$ & 10 & 10 & 20 & 20 & 25 & 150 \\
$p^{\max}_{i}$ & 55 & 55 & 130 & 130 & 165 & 455  \\
$A_i$ & 660 & 670 & 700 & 680 & 450 & 970  \\
$B_i$ & 25.92 & 27.76 & 16.6 & 16.5 & 19.7 & 17.26  \\
$C_i$ & 0.00413 & 0.00173 & 0.002 & 0.00211 & 0.00398 & 0.00031 \\
$D_i$ & 13 & 8 & 12 & 16 & 10 & 12  \\
$E_i$ & 15 & 11 & 13 & 20 & 12 & 15  \\
$F_i$ & 1 & 2 & 3 & 1 & 2 & 2  \\
$T^{down}_{i}$ & 2 & 4 & 1 & 3 & 1 & 2  \\
$T^{up}_{i}$ & 1 & 3 & 2 & 1 & 1 & 2 \\
$r^{down}_{i}$ & 25 & 10 & 30 & 50 & 35 & 60 \\
$r^{up}_{i}$ & 80 & 20 & 20 & 40 & 35 & 50 \\ \hline
i  & 7 & 8 & 9 & 10 & 11 & 12 \\ \hline
$p_{\min,i}$ &  25 & 10 & 150 & 20 & 50 &120\\
$p_{\max,i}$ &  85 & 55 & 455 & 80 & 185 & 370 \\
$A_i$ & 480 & 665 & 1000 & 370 & 490 & 735 \\
$B_i$ &27.74 & 27.27 & 16.19 & 22.26 & 18.5 & 24.9 \\
$C_i$ &0.0079 & 0.00222 & 0.00048 & 0.00712 & 0.0074 & 0.00154\\
$D_i$ & 17 & 12 & 7 & 15 & 15 & 9 \\
$E_i$ & 20 & 14 & 12 & 18 & 18 & 12 \\
$F_i$ &1 & 2 & 1 & 3 & 3 & 1 \\
$T^{down}_{i}$  & 1 & 2 & 1 & 3 & 2 & 2 \\
$T^{up}_{i}$ & 1 & 4 & 1 & 1 & 2 & 3\\
$r^{dn}_{i}$ & 70 & 100 & 80 & 40 & 40 & 80\\
$r^{up}_{i}$  & 15 & 80 & 50 & 30 & 70 & 60\\ \hline
\multicolumn{1}{|c|}{t} & \multicolumn{2}{|c|}{1}  & \multicolumn{2}{|c|}{2} & \multicolumn{2}{|c|}{3} \\ \hline
\multicolumn{1}{|c|}{$L_t$} & \multicolumn{2}{|c|}{1500} & \multicolumn{2}{|c|}{1350} & \multicolumn{2}{|c|}{1450} \\
\multicolumn{1}{|c|}{spinning reserve} & \multicolumn{2}{|c|}{20} & \multicolumn{2}{|c|}{10} & \multicolumn{2}{|c|}{30}\\ \hline \hline
\end{tabular}
\caption{Instance UC\_12a}
\label{tabela:prob10}
\end{table*}
\end{center}

\begin{center}
\begin{table*}[htbp]
\centering
\renewcommand{\arraystretch}{1.2} 
\setlength{\tabcolsep}{10pt}     
\begin{tabular}{|c|c|c|c|c|c|c|}
\hline \hline
i & 1 & 2 & 3 & 4 & 5 & 6  \\ \hline
$p^{\min}_{i}$ &  170 & 20 & 85 & 155 & 195 & 200  \\
$p^{\max}_{i}$ &   355 & 55 & 400 & 360 & 430 & 465   \\
$A_i$ & 960 & 470 & 560 & 400 & 600 & 1000   \\
$B_i$ & 20.4 & 29.8 & 28.5 & 15.9 & 27.9 & 17.2  \\
$C_i$ &  0.00287 & 0.00788 & 0.00646 & 0.00057 & 0.0026 & 0.00584 \\
$D_i$ & 13 & 8 & 12 & 16 & 10 & 12 \\
$E_i$ & 15 & 11 & 13 & 20 & 12 & 15 \\
$F_i$ &  1 & 2 & 3 & 1 & 2 & 2\\
$T^{down}_{i}$ & 2 & 1 & 3 & 2 & 1 & 1   \\
$T^{up}_{i}$ &   1 & 2 & 1 & 2 & 3 & 1 \\
$r^{dn}_{i}$ &  75 & 60 & 40 & 35 & 85 & 70  \\
$r^{up}_{i}$ &  40 & 30 & 50 & 70 & 30 & 40 \\ \hline
i  & 7 & 8 & 9 & 10 & 11 & 12 \\ \hline
$p_{\min,i}$ &  100 & 65 & 15 & 160 & 30 & 60\\
$p_{\max,i}$ &  275 & 305 & 70 & 320 & 220 &  470 \\
$A_i$ & 900 & 910 & 830 & 750 & 860 & 980 \\
$B_i$ & 17.7 & 27.3 & 21.3 & 24.4 & 28.9 & 21.9 \\
$C_i$ & 0.00199 & 0.00454 & 0.0027 & 0.0015 & 0.0026 & 0.00109\\
$D_i$ &  17 & 12 & 7 & 15 & 15 & 9 \\
$E_i$ &  20 & 14 & 12 & 18 & 18 & 12 \\
$F_i$ & 1 & 2 & 1 & 3 & 3 & 1\\
$T^{down}_{i}$  &  2 & 1 & 3 & 2 & 1 & 1\\
$T^{up}_{i}$ &  2 & 1 & 2 & 1 & 2 & 2\\
$r^{down}_{i}$ & 75 &  50 & 85 & 30 & 80 & 65\\
$r^{up}_{i}$  &  70 & 80 & 60 & 100 & 50 & 70\\ \hline 
\multicolumn{1}{|c|}{t} & \multicolumn{2}{|c|}{1} & \multicolumn{2}{|c|}{2} & \multicolumn{2}{|c|}{3} \\ \hline
\multicolumn{1}{|c|}{$L_t$} & \multicolumn{2}{|c|}{2000} & \multicolumn{2}{|c|}{2200} & \multicolumn{2}{|c|}{2500} \\
\multicolumn{1}{|c|}{spinning reserve} & \multicolumn{2}{|c|}{50} & \multicolumn{2}{|c|}{20} & \multicolumn{2}{|c|}{40}\\ \hline \hline
\end{tabular}
\caption{Instance UC\_12b}
\label{tabela:prob11}
\end{table*}
\end{center}

\begin{figure*} [htbp]
\vspace{270pt}
\end{figure*}

\begin{figure*} [htbp]
\section{ UC solutions} 
\end{figure*}

\begin{center}
\begin{table*}[htbp]
\centering
\renewcommand{\arraystretch}{1.2} 
\setlength{\tabcolsep}{10pt}     
\begin{tabular}{|c|c|c|c|c|}
\hline \hline
 Instance & t & Units & Powers & Cost    \\ \hline
  &1 & 0101 & 0,100,0,250 & \multirow{3} {*}{28282.1}   \\
 $\text{UC\_4a}$ & 2 & 0101 & 0,85,0,215  &  \\
  & 3 & 0111 & 0,100,85,315 &  \\ \hline
  & 1 & 1011& 455,0,165,30 &  \multirow{3} {*}{31988.52 } \\
 $\text{UC\_4b}$ & 2 & 1010 &  395,0,135,0 &  \\
 & 3& 1010  & 345,0,105,0 &  \\ \hline
  & 1& 0001010010 & 0,0,0,90,0,355,0,0,455,0 & \multirow{3} {*}{63541.3}  \\
 $\text{UC\_10a}$ & 2&0001010010 & 0,0,0,130,0,415,0,0,455,0 &  \\
 & 3 &0011110010 & 0,0,130,130,130,455,0,0,455,0&  \\ \hline
  & 1& 0011110011 & 0,0,130,130,130,430,0,0,455,25 &  \multirow{3} {*}{79621.3}  \\
 $\text{UC\_10b}$ & 2 & 0011110011 &0,0,130,130,165,455,0,0,455,65& \\
  & 3 &0011010011 & 0,0,130,125,0,455,0,0,455,35 &  \\ \hline
  & 1& 001111001010 & 0,0,130,130,145,455,0,0,455,0,185,0 &  \multirow{3} {*}{ 88046.7}  \\
 $\text{UC\_12a}$ & 2 & 001111001010 & 10,10,130,130,110,410,0,0,455,0,115,0  &   \\
 & 3 &001111001010& 10,10,130,130,130,455,0,0,455,0,150,0  &    \\ \hline
  & 1 & 100101100101& 346,0,0,360,0,444,275,0,0,175,0,400  &  \multirow{3} {*}{ 154514 }  \\
 $\text{UC\_12b}$ & 2 & 100101100101 & 355,0,0,360,0,465,275,0,0,275,0,470  &   \\
&3 & 100111100101&  355,0,0,360,255,465,275,0,0,320,0,470 &    \\ \hline \hline
\end{tabular}
\caption{Best classical solution for the UC intances considered}
\label{tabela:best_class_10}
\end{table*}
\end{center}
\vspace{-100pt}

\begin{center}
\begin{table*} [htbp]
\centering
\renewcommand{\arraystretch}{1.2} 
\setlength{\tabcolsep}{10pt}     
\begin{tabular}{|c|c|c|c|c|}
\hline \hline
 Instance & t & Units & Powers & Cost    \\ \hline
 & 1&0001 & 0, 0, 0, 350 & \multirow{3} {*}{29279.2}   \\
 $\text{UC\_4a}$ & 2&0001 & 0,0,0, 300  &  \\
 & 3 &0101 & 0, 100, 0, 400 &  \\ \hline
& 1&1011& 454.8, 0, 165, 30.2&  \multirow{3} {*}{32370} \\
 $\text{UC\_4b}$ & 2 &1100 & 455.0, 75.0 0, 0 &  \\
  & 3& 1100 & 390.0, 60.0, 0, 0 &  \\ \hline
  & 1 &1000010010 & 21.0, 0, 0, 0, 0, 424.4, 0, 0, 454.6, 0 & \multirow{3} {*}{66771.8}  \\
 $\text{UC\_10a}$ & 2 & 1000011010 & 46, 0, 0, 0, 0, 455, 44, 0, 455, 0 &  \\
  & 3& 1010111010 & 55, 0, 130, 0, 165, 455, 40, 0, 455, 0&  \\ \hline
  & 1 &0011110010 & 0, 0, 130, 129.2, 138.3, 447.7, 0, 0, 454.8, 0 &  
 \multirow{3} {*}{80166.6}  \\
 $\text{UC\_10b}$ & 2& 0011111010 & 0, 0, 130, 130, 165, 455, 65, 0, 455, 0 & \\
  & 3 &0011011010 & 0, 0, 126.1, 125.7, 0, 448.2, 50, 0, 450, 0 &  \\ \hline
 & 1& 000111001001 & 0, 0, 0, 130, 165, 455, 0, 0, 455, 0, 0, 295 &  \multirow{3} {*}{93148.7}  \\
 $\text{UC\_12a}$ & 2&  000011001001& 0, 0, 0, 0, 165, 455, 0, 0, 455, 0, 0, 275 &   \\
  & 3& 000011101001 & 0, 0, 0, 0, 165, 455, 25, 0, 455, 0, 0, 350 &    \\ \hline
  & 1 & 101101101011 & 355, 0, 313.6, 360, 0, 465, 275, 0, 70, 0, 191.4, 470 &  \multirow{3} {*}{162594.4}  \\
$\text{UC\_12b}$ & 2  & 110101101011 & 355, 42.6, 0, 360, 0, 465, 275, 0, 70, 0, 162.4, 470.0 &   \\
 & 3& 010111000101 & 0, 32.2, 0, 360, 352.8, 465, 0, 0,0, 320,0, 470 &    \\ \hline \hline
\end{tabular}
\caption{Standard QAOA results for UC intances 10-15.}
\label{tabela:results_std}
\end{table*}
\end{center}

\begin{center}
\begin{table*} [htbp]
\centering
\renewcommand{\arraystretch}{1.2} 
\setlength{\tabcolsep}{10pt}     
\begin{tabular}{|c|c|c|c|c|}
\hline  \hline
 Instance & t & Units & Powers & Cost    \\ \hline
  & 1&0001 & 0, 0, 0, 350 & \multirow{3} {*}{29279.2}   \\
 $\text{UC\_4a}$ & 2&0001 & 0,0,0, 300  &  \\
 & 3 &0101 & 0, 100, 0, 400 &  \\ \hline
& 1&1011& 454.8, 0, 165, 30.2&  \multirow{3} {*}{32370} \\
 $\text{UC\_4b}$ & 2 &1100 & 455.0, 75.0 0, 0 &  \\
  & 3& 1100 & 390.0, 60.0, 0, 0 &  \\ \hline
  & 1 &1000010010 & 21.0, 0, 0, 0, 0, 424.4, 0, 0, 454.6, 0 & \multirow{3} {*}{66771.8}  \\
 $\text{UC\_10a}$ & 2 & 1000011010 & 46, 0, 0, 0, 0, 455, 44, 0, 455, 0 &  \\
  & 3& 1010111010 & 55, 0, 130, 0, 165, 455, 40, 0, 455, 0&  \\ \hline
  & 1 &0011110010 & 0, 0, 130, 129.2, 138.3, 447.7, 0, 0, 454.8, 0 &  
 \multirow{3} {*}{80166.6}  \\
 $\text{UC\_10b}$ & 2& 0011111010 & 0, 0, 130, 130, 165, 455, 65, 0, 455, 0 & \\
  & 3 &0011011010 & 0, 0, 126.1, 125.7, 0, 448.2, 50, 0, 450, 0 &  \\ \hline
  & 1 & 001111001010 & 0, 0, 129.9, 130, 161, 455, 0, 0, 454.6, 0, 169.5, 0 &  \multirow{3} {*}{89277.7}  \\
$\text{UC\_12a}$ & 2 & 001011001010 & 0, 0, 130, 0, 145.8, 455, 0, 0, 455, 0, 164.2, 0 &   \\
  & 3 & 101011011010 & 34.3, 0, 130, 0, 165, 455, 0, 25.7, 455, 0, 185, 0 &    \\ \hline
  & 1 & 110101101001& 355, 20, 0, 360, 0, 465, 275, 0, 70, 0, 0, 455 &  \multirow{3} {*}{158406.1}  \\
$\text{UC\_12b}$ & 2  & 110101101011 & 355, 35.4, 0, 360, 0, 465, 275, 0, 70, 0, 169.6, 470.0 &   \\
 & 3& 100111100011 & 355, 0, 0, 360, 380.2, 465, 275, 0, 0, 0, 194.8, 470 &    \\ \hline \hline
\end{tabular}
\caption{Warm-Start QAOA results for UC intances 1-6.}
\label{tabela:results_ws}
\end{table*}
\end{center}

\vspace{200pt}


\end{document}